\documentclass[journal]{IEEEtran}
\usepackage{lipsum} 
\usepackage[tight,footnotesize]{subfigure}
\usepackage{enumitem} 
\usepackage{float}
\usepackage{multicol,multienum}
\usepackage{breqn}
\usepackage{stfloats}
\usepackage{multirow}
\usepackage{diagbox}
\usepackage{hyperref}
\usepackage{bm}
\usepackage{cite}
\usepackage{graphicx}
\usepackage{tabularx}
\usepackage[ruled,vlined]{algorithm2e} 
\usepackage{booktabs}

\usepackage[tight,footnotesize]{subfigure}
\usepackage{url}
\usepackage{doi}
\usepackage{amssymb}
\usepackage{color}
\ifCLASSINFOpdf
 
\else
 
\fi

\begin{document}

\title{Federated Learning for 6G Communications: Challenges, Methods, and Future Directions}


\author{Yi~Liu,
    Xingliang~Yuan,~\IEEEmembership{Member,~IEEE,}
    Zehui~Xiong,~ Jiawen~Kang,~ Xiaofei Wang,~\IEEEmembership{Senior Member,~IEEE,}
    Dusit Niyato,~\IEEEmembership{Felow,~IEEE} 
\thanks{Yi Liu is with School of Data Science of Technology, Heilongjiang University, Harbin, China (97liuyi@ieee.org).}
\thanks{Xingliang Yuan is with Faculty of Information Technology, Monash University, Clayton VIC, 3800, Australia (e-mail: xingliang.yuan@monash.edu).}
\thanks{Zehui Xiong and Dusit Niyato are with School of Computer Science and Engineering, Nanyang Technological University, Singapore.  (e-mail: zxiong002@e.ntu.edu.sg, dniyato@ntu.edu.sg).}
\thanks{Xiaofei Wang is with the College of Intelligence and Computing, Tianjin University, Tianjin, China. (e-mail: xiaofeiwang@tju.edu.cn).}
\thanks{Jiawen Kang is with Energy Research Institute, Nanyang Technological University, Singapore (e-mail: kavinkang@ntu.edu.sg).}
}

\markboth{China Communications}%
{Shell \MakeLowercase{\textit{et al.}}: Bare Demo of IEEEtran.cls for IEEE Journals}

\maketitle

\begin{abstract}
As the 5G communication networks are being widely deployed worldwide,  both industry and academia have started to move beyond 5G and explore  6G communications. It is generally believed that 6G will be established on ubiquitous Artificial Intelligence (AI) to achieve data-driven Machine Learning (ML) solutions in heterogeneous and massive-scale networks. However, traditional  ML techniques require centralized data collection and processing by a central server, which is becoming a bottleneck of large-scale implementation in daily life due to significantly increasing privacy concerns. Federated learning, as an emerging distributed AI approach with privacy preservation nature, is particularly attractive for various wireless applications, especially being treated as one of the vital solutions to achieve ubiquitous AI in 6G. In this article, we first introduce the integration of 6G and federated learning and provide potential federated learning applications for 6G. We then describe key technical challenges, the corresponding federated learning methods, and open problems for future research on federated learning in the context of 6G communications.
\end{abstract}

\begin{IEEEkeywords}
6G communication, federated learning, security and privacy protection
\end{IEEEkeywords}

\IEEEpeerreviewmaketitle
\IEEEpeerreviewmaketitle
\section{Introduction}

\begin{figure*}[!t]
	\centering
	\includegraphics[width=1\linewidth]{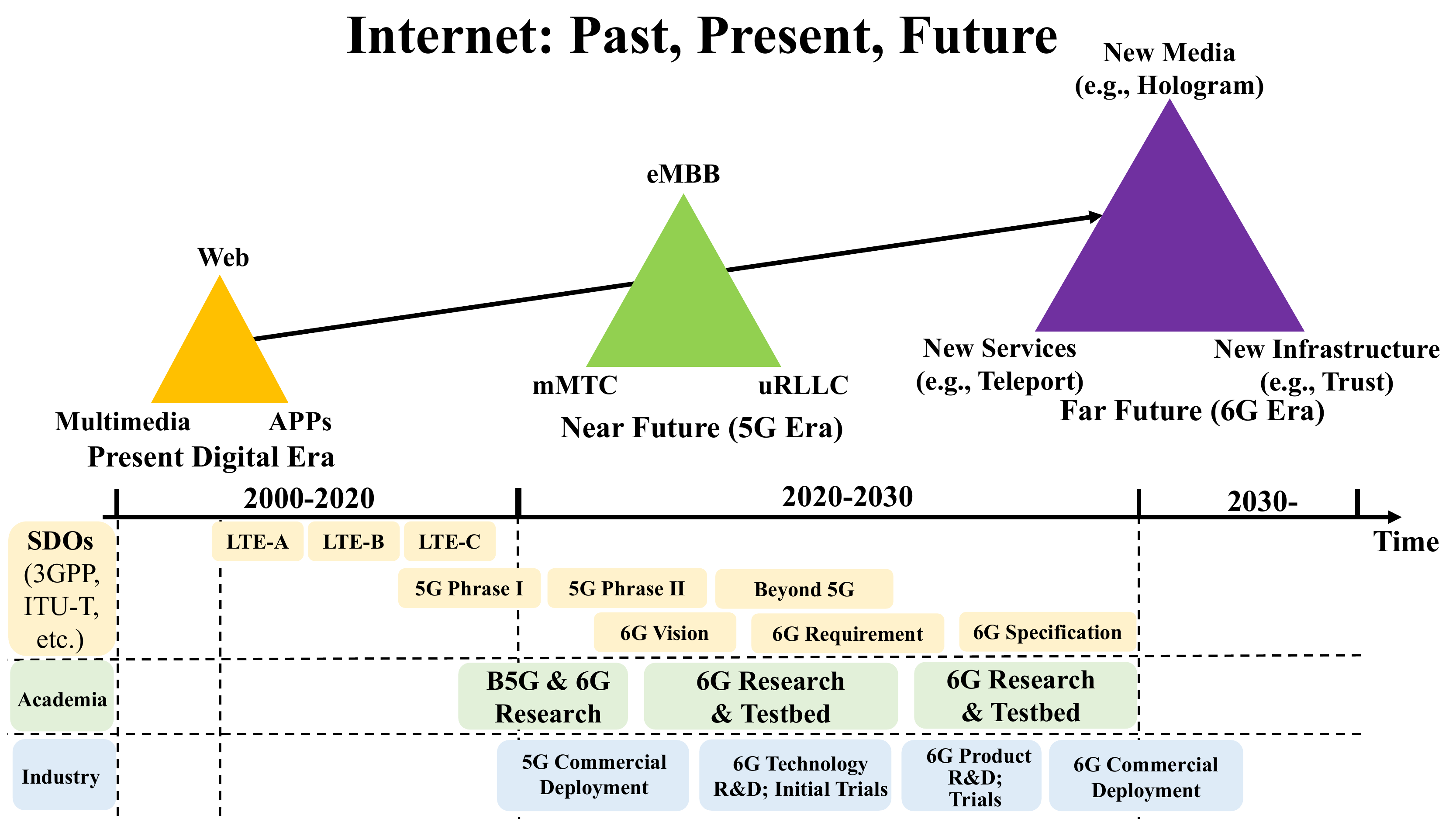}
	\caption{Key services and roadmap for 6G \cite{ref-1,ref-2,ref-3,ref-4}.}
	\label{fig-1}
\end{figure*}

\IEEEPARstart{T}{he} rapid development of wireless communication techniques with numerous technological innovations for decades has greatly improved people's lives and promoting the development of the industry. As shown in Fig. \ref{fig-1}, the current Fourth-generation (4G) LTE network has created an era of mobile internet with Web search services, multimedia services, and APPs as core functions \cite{ref-1}. The upcoming Fifth-generation (5G) system is designed to support a wider range of services, such as Augmented Reality/Virtual Reality (AR/VR), large-scale Internet of Things (IoT), and autonomous driving \cite{ref-2}. Specifically, the 5G system includes three technical characteristics: enhanced Mobile BroadBand (eMBB), massive Machine-Type-Communications (mMTC), ultra-Reliable Low-Latency Communications (uRLLC).  Since 5G brings unprecedented benefits to humans and is being actively deployed around the world,  both industry and academia have begun to move towards the next generation of wireless technology, i.e., Sixth-generation (6G) \cite{ref-1, ref-2}.

The 5G system represents a new wireless communication paradigm that adopts a service-based architecture (SBA) instead of a communication-oriented architecture (COA) to achieve ``connected things''.  In contrast to previous generations, 6G with transformative technologies will revolutionize the development of wireless communication from ``connected things" to ``connected intelligence" \cite{ref-1}.  Specifically, 6G will revolutionize technology in three areas: new media, new services, and new infrastructures. It is expected that the 6G system will adopt advanced artificial intelligence (AI) technologies in these fields, and promptly and efficiently collect, transmit, and learn data anytime, anywhere to generate a large number of innovative applications and intelligent services \cite{ref-4}. In particular, ubiquitous AI will empower the promising 6G, a hyper-flexible architecture that brings human-centric development concepts to all aspects of network systems, instead of data-centric, machine-centric, and application-centric \cite{ref-3}. Therefore, 6G communications have higher-level security and stronger privacy protection requirements.

{However, traditional Machine Learning (ML) empowered frameworks based on a central server are suffering from critical privacy and security challenges, e.g., single point of failure, which is not able to enable ubiquitous and secure AI for 6G. Moreover, due to large overhead caused by centralized data aggregation and processing, traditional centralized ML schemes might not be suitable for ubiquitous ML \cite{niknam2019federated}. Thereby, decentralized ML solutions, in which all private data is kept in training devices locally, are becoming increasingly essential for 6G. Recently, Federated Learning (FL) as an emerging decentralized ML solution has attracted particular attention from academia and industry \cite{ref-2,konevcny2016federated}.  
In FL, participating devices collaboratively train a shared model through their local data, and thus only upload model updates instead of raw data to centralized parameter servers \cite{kang2019incentive}. }
	
Although FL brings high potential for AI-empowered 6G and significantly improves privacy-sensitive applications with 6G communication, FL is still in the early stages of development and is facing new challenges in 6G scenarios. In this paper,  we first introduce the core challenges of FL in 6G communications including (i) large communication cost due to multiple communication rounds for model updates and aggregation \cite{ref-13}; (ii) security problems caused by heterogeneous and diverse participating entities,  e.g., poisoning attacks and backdoor attacks \cite{kang2019incentive, ref-10}; (iii) privacy problems resulted from gradient leakage attacks and membership inference attacks \cite{ref-15}; (iv) model training and inference efficiency problems among massive-scale 6G networks. We then propose advanced federated learning methods to address the above challenges from different perspectives. Finally, we describe the open research topics and future directions of FL in 6G communications. 

\section{Preliminaries and Overview}
\subsection{{Key 6G Requirements and Use Cases}}
Unlike 5G communications, 6G has prominent features to ensure ubiquitous, seamless, intelligent, high-performance connectivity and networking with security and privacy protection. More specifically, we will introduce the performance requirements of 6G communications as follows. 
\subsubsection{\textbf{High Performance Networking}}
It is commonly believed that 6G is a complex networking system with many heterogeneous space-air-ground-underwater communication networks \cite{ref-2, ref-3}. The three-dimensional super-connectivity networks provide worldwide connectivity and integrated networking to enable different types of network services and dense coverage through sub-networks and sub-systems, e.g., satellite communication networks, underwater-land communications. With the help of massive-scale heterogeneous networks, 6G communications can achieve up to 1 Tbps data rate per user, ultra-low end-to-end delay, superior end-to-end reliability, and high energy efficiency networking \cite{ref-2}. Compared with 5G communications, 6G communications support networking and connecting the majority not only in dense areas but also the less dense areas, such as the underwater environment,  in an efficient and low overhead manner  \cite{ref-3}.  6G communications employ novel communication networks to support highly diversified data, e.g., audio, video, AR/VR data, which reaches new communication experience with virtual networking existence and involvement anywhere  \cite{ref-2}.
\subsubsection{\textbf{Higher Energy Efficiency}}
In the 6G era, there exist higher energy efficiency requirements for wireless devices with charging constraints and battery life limitations. Therefore, long battery life and low energy consumption are two popular research topics for 6G communications.  To address the energy problems of wireless devices, especially smartphones, existing studies have proposed energy harvesting technology,  wireless power transfer technology, and green communication to improve energy efficiency and extend the working time of wireless devices \cite{ref-3}.
Especially, the wireless devices can harvest energy from ambient radio-frequency, solar, geothermal energy, and wind energy by using different energy harvesting technologies, which can prolong the battery life. Similarly, the wireless devices with wireless charging equipment can obtain energy supplement from dense network infrastructures or mobile charging stations, e.g., Unmanned Aerial Vehicle (UAV), Electric Vehicles (EVs), through wireless power transfer technology.

{Recently, to address energy problems for wireless devices, and emerging technology named symbiotic radio (SR) is introduced to integrate passive backscatter devices with an active transmission system \cite{ref-42,ref-43}. A typical example of SR is ambient backscatter communication, that enables network devices to utilize ambient RF signals to transmit information without requiring active RF transmission, making battery-free communication possible \cite{ref-42,ref-44}. Smart energy management is another promising mechanism with the goal of dynamically optimizing the balance between energy demand and supply \cite{ref-42}.}

For green communication techniques, AI-based solutions are quite important to optimize energy usage and energy scheduling for wireless devices in a dynamic environment and complex optimization goals. Advanced machine learning techniques, such as deep reinforcement learning, can be utilized to optimize the computation task offloading decision of a wireless device, and also make the best scheduling solution of working and sleeping time, which can lower energy consumption and enhance energy efficiency. The AI-based solutions can be also applied in multi-hop information routing in cooperative relay communication and communication infrastructure deployment in network-densification 6G scenarios, which significantly reduces the transmit power of the wireless devices without long propagation distance thus enabling high-efficiency communication \cite{ref-2,ref-3}.

\subsubsection{\textbf{High Security and Privacy}}
Existing research mainly focuses on network throughput, reliability, and delay in 4G and 5G communications \cite{ref-3}. However, in the past few decades, wireless communication security and privacy issues have been ignored to some extent. Since data security and privacy issues are closely related to users' lives, protecting data security and privacy has become a very important part of human-centric 6G communications. Meanwhile, communication/data service providers legally collect a large amount of user information, which leads to frequent leakage of privacy data. In order to solve this problem, it is envisaged that FL techniuqes can be used to achieve privacy-enhanced deep learning in 6G networks.

\subsubsection{\textbf{High Intelligence}}
The high intelligence of 6G will be beneficial to provide users with high-quality, personalized, and intelligent services. High-intelligent 6G includes operational intelligence, application intelligence, and service intelligence as follows.

\begin{itemize}
\item  \textbf{Operational Intelligence.} Traditional network operations involve a series of resource optimization and multi-objective performance optimization problems \cite{ref-1}. In order to achieve a satisfactory level of network operation, optimization methods based on game theory, contract theory, etc. are widely used. However, these optimization theories may not obtain the optimal solution in large-scale time-varying variables and multi-objective scenarios. With the development of deep learning technologies, the above can be solved by using advanced machine learning technologies. On the other hand, the emergence of federated learning has transformed the multi-objective linear optimization problems into a nonlinear optimization problem, thus finding out the best solution for complex and time-varying decisions in operational intelligence  \cite{ref-2}.
\item  \textbf{Application Intelligence.} At present, applications related to 5G networks are gradually becoming intelligent. For 6G networks, intelligent applications are one baseline of application requirements \cite{ref-10,ref-18}. FL empowered wireless communication technologies to enable devices to connect with 6G networks to run a variety of intelligent applications. For example, in the future, users may need intelligent voice assistants to complete their daily schedules \cite{ref-12}. The  6G network ubiquitous AI will provide users with highly intelligent applications.
\item \textbf{Service Intelligence} Furthermore, as a human-centric network, the high intelligence of the 6G network will provide intelligent services in a satisfactory and personalized manner \cite{ref-1,ref-2,ref-3}. For example, FL provides users with personalized healthcare services, personalized recommendation services, and personalized intelligent voice services in a distributed learning manner. In the future, intelligent services will be tightly integrated with the 6G networks \cite{ref-3}.
\end{itemize}

{\subsubsection{\textbf{Increased Device Density}} Compared with 5G, the 6G has much higher transmission rates and shorter delay, greater device density, and the integration of Artificial Intelligence (AI). With the increased device density and explosive increasing data traffic, it is more and more important to solve the network capacity challenges. One of the potential solutions is to provide increasingly more but smaller radio cells that can transmit data quickly and energy-efficiently.  These cells are required to be connected as seamlessly as possible to the fiber-optic core networks via high-performance transmission links. An important goal is to connect these wireless transmission links directly to fiber-optic networks without complex electronics. Thus the fiber optic networks can provide extremely high transmission capacity and reliability for massive devices with insignificant latency through flexible and ubiquitous wireless networks \cite{ref-41}.}

{\subsubsection{\textbf{Green Communication}} It is significant for green communication to make good decisions for optimizing resource utilization and communication efficiency. In 6G communication scenarios, due to massive network traffic, innumerable network devices, and dynamic network environments, there exist more and more complex resource optimization problems, e.g., green communication optimization and offloading decision, that traditional mathematical programming techniques and optimization solutions cannot tackle.  Recently, data science and AI-based optimization have also largely been used to solve problems related to resource optimization, task assignment in distributed systems, because of its advantages of data-driven decision, dynamic flexibility, and self-adjustment. }
\subsection{Typical Use Scenarios}
\begin{figure*}[!t]
	\centering
	\includegraphics[width=1\linewidth]{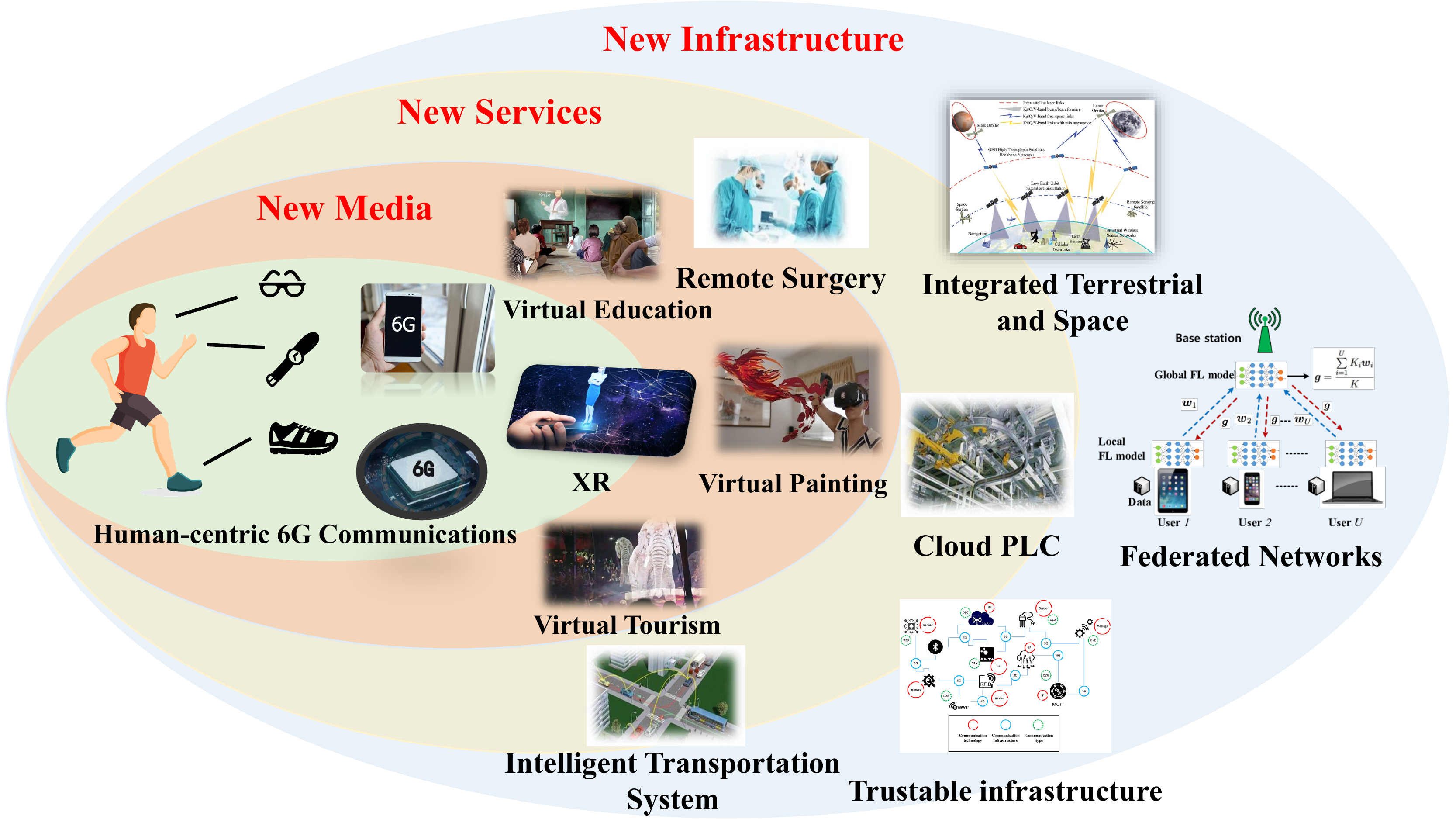}
	\caption{{Key services and roadmap for 6G \cite{ref-38,ref-39,ref-40}.}}
	\label{fig-6}
\end{figure*}

Compared with previous generations, the 5G service model has been transformed into a service-based architecture, and its user cases include: enhanced Mobile Broad Band (eMBB), massive Machine-Type-Communications (mMTC), and ultra-Reliable Low-Latency Communications (uRLLC). As shown in Fig. \ref{fig-6}, driven by Industry 5.0 and deep learning technologies, 6G will provide the following new service types:

\begin{itemize}
\item \textbf{New Media.} With the rapid development of wireless network communication technologies, it can be expected that the form of information interaction will gradually evolve from AR/VR to high fidelity extended reality (XR) interaction after 10 years, and even realizing wireless holographic communication \cite{ref-1,ref-2,ref-3,ref-4}. Users can enjoy the new services brought by holographic communication and holographic display anytime and anywhere, such as virtual education, virtual tourism, virtual sports, virtual painting, virtual concerts, and other fully immersive holographic experiences.

\item  \textbf{New Services.} According to ITU-T's 6G communication technology white paper, beyond and high-precision teleport technology will provide users with a variety of new services \cite{ref-1,ref-2,ref-3,ref-4,ref-5}. Holographic teleport, quantum communication, visible light communication (VLC), and other communication technologies have subverted the traditional service model. For example, industries such as remote surgery \cite{ref-33}, cloud PLC \cite{ref-34}, and intelligent transportation systems \cite{ref-35} will be empowered by the new service model to provide users with better services. The goal of these new technologies is to provide high-precision services, deterministic service, and best-guaranteed services.

\item  \textbf{New Infrastructure.} With the development of deep learning technologies, the 6G communication system has spawned many emerging infrastructures such as Integrated Terrestrial and Space \cite{ref-34}, federated learning networks \cite{ref-18}, decentralized infrastructures \cite{ref-1}, and trustable infrastructure \cite{ref-3}. In particular, the FL network benefits from the high bandwidth and low latency of the 6G network, which has brought many emerging intelligent applications to cities, factories, and people.
\end{itemize}

\subsection{Federated Learning}
In this subsection, we introduce an FL-based distributed learning architecture in 6G.  In this architecture, a large number of decentralized devices associated with different services can collaboratively train a shared global model (e.g., anomaly detection, recommendation system, next-word prediction, etc.) by using locally collected datasets.

\begin{figure*}[!t]
	\centering
	\includegraphics[width=1\linewidth]{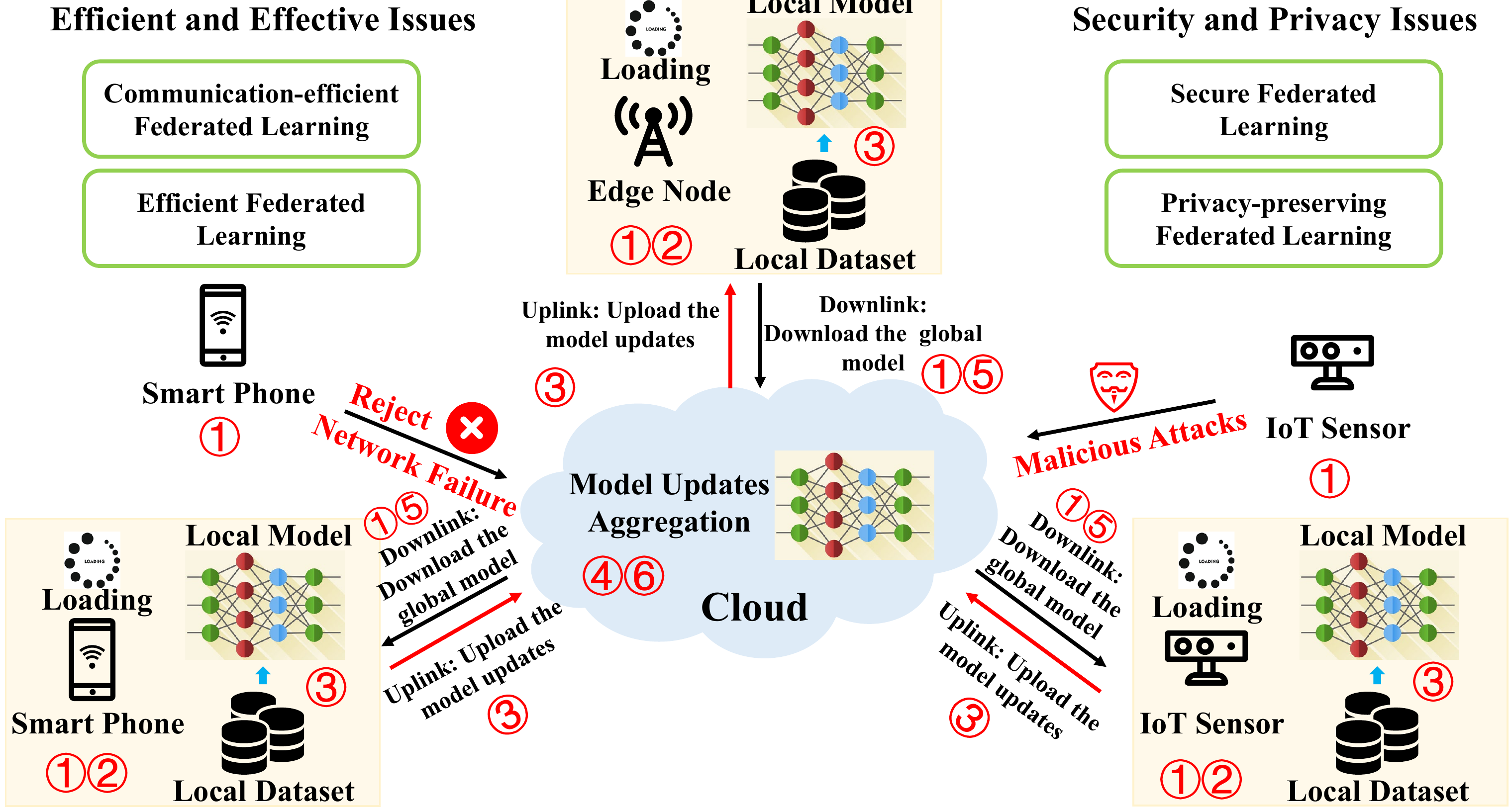}
	\caption{An overview of federated learning process in 6G \cite{ref-32}.}
	\label{fig-4}
\end{figure*}

As shown in Fig. \ref{fig-4}, the procedure of FL-based architecture is divided into three phases: the initialization, the training, and the aggregation phase. In the initialization phase, a device will evaluate its service requests, needs, and connection conditions, and decides whether to register with the nearest cloud to join the training of the shared global model via a wired or wireless connection (e.g., 6G). Then, the cloud acting as task publisher will randomly select a subset of devices from the registered devices to participate in this round of training, and reject the remaining registered devices. {The cloud will also send initialized or pre-trained global model  ${\omega _{t}}$ to each selected device (steps \textcircled{\scriptsize 1}, \textcircled{\scriptsize 2}). In the training phase, each selected device trains global model $\omega _t^k \leftarrow {\omega _t}$ by using local dataset to obtain the updated global model $\omega _{t+1}^{k}$ in {each iteration}. In particular, for the $k$-th device {($k \in \{ 1,2, \ldots ,K\}$)}, the loss function needs to be optimized as follows: $	\arg \mathop {\min }\limits_{\omega  \in \mathbb{R}}{F_k}(\omega ),{F_k}(\omega ) = \frac{1}{{{D_k}}}\sum\nolimits_{i \in {D_k}} {{f_i}} (\omega )$, where $D_k$ denotes the size of local dataset that contains input-output {vector} pairs $(x_i, y_i)$, ${x_i},{y_i} \in \mathbb{R}$, $\omega$ is local model parameter, and ${f_i}(\omega )$ is a local loss function (e.g., ${f_i}(\omega ) = \frac{1}{2}({x_i}^T\omega  - {y_i})$). Each selected device uploads the model updates to the cloud (steps \textcircled{\scriptsize 3}, \textcircled{\scriptsize 4}, \textcircled{\scriptsize 5}). In the aggregation phase, the cloud receives model updates of all selected devices for aggregation to obtain a new global model $\omega_{t + 1}$ for the next iteration, i.e., $	{\omega _{t + 1}} \leftarrow {\omega _t} - \frac{1}{K}\sum\limits_{k = 1}^K {{F_k}(\omega )}$, where $K$ denotes the number of edge nodes.} In the next round, the device selected by the cloud downloads the current latest global model $\omega_{t + 1}$ from the cloud. The device will use the received new global model to update its respective model. In the next round of training, the cloud will randomly select a new device subset and repeat the above process until the trained model converges or meets the stopping criteria (step \textcircled{\scriptsize 6}). 
\section{Core Challenges for Federated Learning in 6G}
In this section, we introduce the core challenges of FL, which are the main bottleneck problem before large-scale deployment of FL in 6G applications. 
\subsection{Challenge 1: Expensive Communication}
Since the FL involves thousands of devices participating during model training, communication is a critical bottleneck for FL being widely used in 6G \cite{ref-18}. Previous studies \cite{konevcny2016federated, kangincentive, ref-13, ref-15,ref-24,ref-25,ref-26,ref-32} has made many efforts to improve the communication efficiency of FL system. Furthermore, it is challenging for FL networks to achieve communication in the FL networks is synchronized with the local calculation of the device \cite{ref-13,ref-14,ref-15}. To make the FL model suitable for 6G networks with massive, heterogeneous devices and networks, it is necessary to develop a communication-efficient method, which can greatly reduce the number of gradients exchanged between the devices and the cloud instead of all gradients information. In order to further reduce communication overhead in this setting, two key aspects need to be considered: (i) reducing the total number of communication rounds, or (ii) reducing the number of gradients in each communication round.
\subsection{Challenge 2: Security Problems}
Since 6G networks can provide ubiquitous services across a wider geographic area, the computing and communication capabilities of each device in the network may vary due to changes in hardware (CPU, GPU), network connectivity (4G, 5G, 6G, WiFi), and energy (battery level). Obviously, the system heterogeneity between the devices will bring some confusion and faults to the FL model and 6G network \cite{ref-10,ref-18}. Additionally, there may be unreliable devices in the FL, which may  cause the Byzantine failure of the system. Similarly, adversaries may launch active learning-based attacks (like poisoning attacks and backdoor attacks) on heterogeneous devices and cause errors in the FL system. The security vulnerabilities of these FL systems greatly exacerbate challenges such as mitigating attacks, tolerance, and faults. Therefore, developing a secure and robust FL must: (i) defend against malicious attacks, (ii) tolerate heterogeneous hardware, and (iii) achieve robust aggregation algorithms.
\subsection{Challenge 3: Privacy Concerns}
Although FL protects the privacy of each device by sharing model updates (e.g., gradients information) instead of the raw data, the private data will still be disclosed during the interaction between the device and the cloud \cite{ref-19}. For example, adversaries will launch \textbf{membership inference} \cite{ref-19} or \textbf{gradient leakage attacks} \cite{ref-7} to steal local training data from the devices. Previous work has focused on using tools such as secure multi-party (SMC) computing or homomorphic encryption (HE) to enhance the privacy of FL, but these methods cannot address the above malicious attacks \cite{ref-18}. SMC and HE can only prevent data breach problems and cannot resist member inference attacks and gradient leak attacks. Therefore, it is very urgent for the FL system to develop new privacy-enhancing techniques to resist or mitigate the aforementioned malicious attacks.
\subsection{Challenge 4: Effective Issues}
Deploying FL models to devices generally involves model training and inference \cite{ref-18}. If the speed of model training and reasoning is relatively slow, users will not be able to experience real-time intelligent services \cite{ref-23}. Therefore, when FL systems are widely deployed in 6G networks, they will encounter the following challenges: (i) the size of the FL model is too large to adapt to a single device; (ii) the FL model training is too slow to meet the delay requirements of the 6G network; (iii) the FL model inference is too slow to satisfy the user' real-time demand. Efficient training and inference are necessary for the perfect integration of FL and 6G networks. However, it is challenging for FL systems to achieve efficient model training and inference in a massive, heterogeneous network.
\section{Advanced Federated Learning Methods For 6G}
To address the aforementioned challenges, we propose advanced federated learning systems through different emerging technologies or methods to enable communication-efficient, secure, and privacy-enhanced federated learning, respectively. 
\subsection{Communication-efficient Federated Learning For 6G}
In 6G, it is challenging for devices that span a larger geographic area to obtain a better global model but with huge communication overhead. The communication overhead will affect the gradient exchange between the devices and the cloud, thus affecting the model aggregation at the cloud. Therefore, we need to find a more efficient way to achieve FL training. In this subsection, we will explain communication-efficient FL from the perspectives of system-level and algorithm level, which promotes a wider-range FL deployment and usage for 6G communications.
\subsubsection{\textbf{Communication-efficient FL: System Level}}\label{asy}
\begin{figure}[!t]
	\centering
	\includegraphics[width=0.9\linewidth]{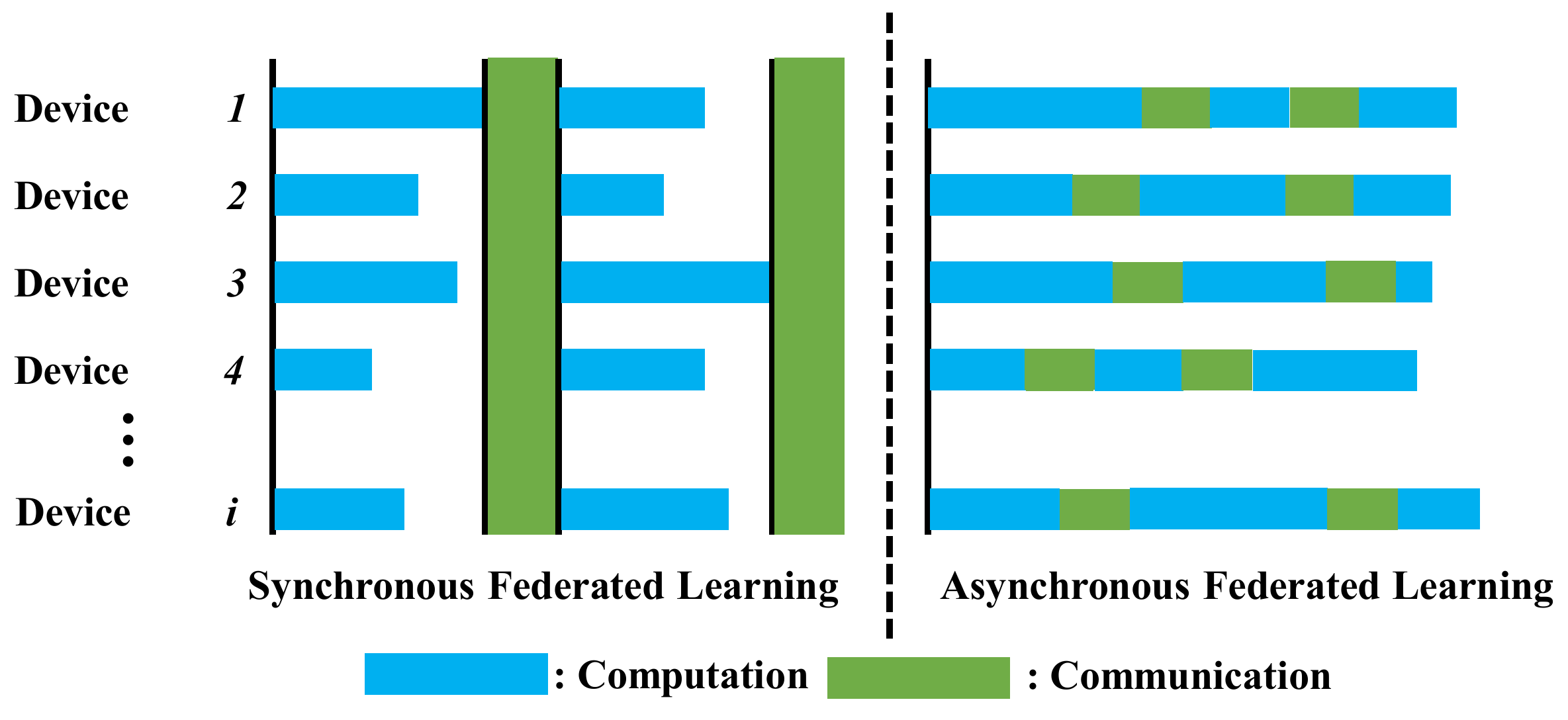}
	\caption{The overview of the synchronous and asynchronous FL.}
	\label{fig-5}
\end{figure}

From a system perspective, data distribution (e.g., non-independent and identical distribution), device distribution (e.g., heterogeneous devices across regions and networks), computation methods (e.g., decentralized and centralized), and communication mechanisms (e.g., synchronous and asynchronous scheme) have different impacts on communication efficiency in different application scenarios \cite{ref-13}. 
\begin{itemize}
    \item  \textbf{Asynchronous FL System:} As shown in Fig. \ref{fig-5}, AFLS can reduce the {computation time} of the devices by asynchronously aggregating the model updates, thereby improving the communication efficiency of {FL}. {Let $\kappa  = \frac{{Comm}}{{Comp + Comm}}$, where $\kappa$ represents communication efficiency, $Comm$ is the communication time, and $Comp$ is the computation time. It can be seen from Fig. \ref{fig-5} that the $Comp$ of asynchronous model update scheme is shorter than that of synchronous one, so the communication efficiency $\kappa$ of the asynchronous model update scheme is higher than that of  the synchronous one.} 
\end{itemize}
\subsubsection{\textbf{Communication-efficient FL: Algorithm Level}}
At the algorithm level, achieving communication-efficient FL can reduce the communication rounds of training a model by accelerating convergence \cite{ref-14} and reduce the communication cost of each round by using gradient compression techniques \cite{ref-15} (e.g., sparsification, quantization, etc.).  More details are described below.
\begin{itemize}
    \item \textbf{Accelerating Model Convergence:} Stochastic gradient descent (SGD) algorithms based on zero-order, first-order, second-order, and federated optimization are used to reduce the number of rounds of model training \cite{ref-13}. Since the federated optimization method can protect the private data on each device, it is very popular with this unique motivation in accelerating training model convergence. 

    \item \textbf{Reducing Communication Overhead:} Gradient sparsification and gradient quantization can greatly reduce the large number of gradients exchanged between the devices and the cloud to achieve communication-efficient FL. Lin \textit{et al.} in \cite{ref-15} proposed a Top-k selection-based gradient compression scheme to improve communication efficiency. In this scheme, the authors can compress the gradient 300 times to reduce the number of gradients without compromising accuracy.
\end{itemize}

\subsection{Secure Federated Learning For 6G}
Due to the wide range of 6G network connections, FL will suffer malicious attacks from heterogeneous networks, heterogeneous devices, and malicious participants during the training process \cite{ref-10}. To alleviate this problem, researchers have proposed many different defense solutions from three perspectives: aggregation algorithm, detection mechanism, and reputation management. 

\subsubsection{\textbf{Robust Aggregation Algorithm}} Aggregation is a very important operation in the FL training process that directly affects the results of model convergence. The motivation of the robust aggregation algorithm is to greatly reduce the impact of low-quality model updates generated by malicious devices (i.e., poisoning attacks) on global model training. Furthermore, this method can make the cloud tolerate Byzantine failures of some devices \cite{ref-36,ref-37}. For example, Ang \textit{et al.} in \cite{ref-11} proposed the regularizer approximation method to reduce the noise interference of heterogeneous devices and heterogeneous networks. 

\subsubsection{\textbf{Robust Detection Mechanism}} Another intuitive idea is to detect malicious devices to prevent them from participating in FL training. Such a mechanism has generally utilized the accuracy of the sub-model generated by the device as an evaluation metric to detect malicious devices. Liu \textit{et al.} in \cite{ref-10} utilized the smart contract techniques in the blockchain to design a malicious device detection mechanism to alleviate the malicious attack problems.

\subsubsection{\textbf{Reliable Reputation Management}} The historical behaviors of the devices can be used as a key indicator to evaluate its reliability and trustworthiness by a metric named reputation. The high reputation value indicates more reliable devices. Inspired by this, establishing a reputation management scheme for device historical behaviors in FL can also prevent malicious devices from damaging the global model. Kang \textit{et al.} in \cite{ref-12} proposed a reputation management scheme to calculate the historical reputation of the devices to achieve a robust FL with high-reputation devices.

\subsection{Privacy-preserving Federated Learning For 6G}
\subsubsection{\textbf{Differentially Privacy}} Differential privacy (DP) \cite{ref-8} techniques are proposed to protect the privacy of gradient information, thereby achieving cloud-level privacy protection. Geyer \textit{et al.} in \cite{ref-9} applied the DP technique in FL system that protects cloud-level privacy. Similarly, in order to protect user-level privacy, the local differential privacy (LDP) techniques achieve this goal by disturbing the gradients uploaded by the devices \cite{ref-19}. However, DP and LDP technologies enhance FL privacy at the expense of model performance. Therefore, there are currently advanced methods that balance privacy and performance as described below.

\subsubsection{\textbf{Deep Net Pruning}} Neural network pruning is a technique of deep learning whose goal is to develop a smaller and more efficient neural network. Recently, Huang \textit{et al.} \cite{ref-20} utilized pruning as an equivalent technique of DP to protect the privacy of the FL system while ensuring the model performance. Such a method creates a new idea of using model pruning to be equivalent to DP techniques, which provides new opportunities for balancing utility and privacy.

\subsubsection{\textbf{Gradient Compression}} The reason why adversaries can infer the local data of the devices is that the gradient information contains rich semantic information \cite{ref-7}. Inspired by the above, an intuitive idea is that the methods that disrupt the distribution of gradient information thus protecting the gradient privacy. Zhu \textit{et al.} in \cite{ref-7} proved that gradient compression can defend against gradient leakage attacks without compromising accuracy and the defense effect is better than that of DP.

\subsection{Effective Federated Learning For 6G}
The long-term goal of human-centric communication services in 6G networks is to  handle user needs in real time. Therefore, it is necessary to achieve efficient FL from training and inference.
\subsubsection{\textbf{Efficient Training}} Efficient training can greatly reduce the training time of mobile devices to achieve efficient FL. The advanced training methods are summarized as follows.

\begin{itemize}
    \item  \textbf{{Federated Parallelization:}} Data and model parallelization are generally used to accelerate model training. Data parallization achieves efficient training by running multiple training samples in parallel \cite{ref-21,ref-22}. Model parallelization accelerates model training by splitting the model over multiple processors \cite{ref-23}.

 \item \textbf{Federated Distillation:} Model distillation adopts transfer learning to utilize the output of a pre-trained complex model (i.e., Teacher model) as a supervised signal to train another simple network, i.e., Student model. Such a way can train student models to improve the efficiency of model training. Jeong \textit{et al.} in \cite{ref-24} proposed federated distillation (FD), an efficient distributed model training algorithm, whose training efficiency is much smaller than the FL benchmark scheme, especially when the model size is large.

\end{itemize}

\subsubsection{\textbf{Efficient Inference:}} The size of the existing FL model is too large to realize real-time inference on the devices. Efficient inference can be achieved in the following ways.

\begin{itemize}
    \item  \textbf{Pruning:} The pruning technique is a model optimization technique that includes removing excess weights in the weight tensor. The compressed neural network not only runs faster but also reduces the computational cost of the training network, which is a critical step in deploying the model to mobile phones or other edge devices. 

  \item  \textbf{Weight Sharing.} Weight sharing reduces the number of model parameters by sharing weights, thereby achieving efficient model inference. The reason is that the fewer the parameters of the model, the smaller the model size. Tran \textit{et al.} in \cite{ref-25} utilized weight sharing approach for wireless networks to improve model inference efficiency. 
  \end{itemize}
  
\section{Open Research Topics And Future Directions}
\subsection{Trustworthy Federated Learning}
\subsubsection{\textbf{Privacy-enhanced Federated Learning}}
Previous work about FL has covered user or cloud-level privacy for all devices in the 6G networks. However, in practice, the previous schemes provide strict privacy restrictions at the expense of accuracy \cite{ref-18}. It is essential for FL to develop privacy-enhanced techniques that do not compromise accuracy to provide strict privacy guarantees because the industry is very concerned about the accuracy of the FL model. To this end, few studies are exploring potential solutions. For example, Huang \textit{et al.} \cite{ref-20} recently proposed a net pruning technique to provide strict privacy guarantees by replacing pruning with DP technique, and also to improve the training efficiency of the model.  It is an interesting and ongoing direction to developing methods that can balance efficiency and privacy restrictions in future work.
\subsubsection{\textbf{Security-enhanced Federated Learning}}
Since the FL systems normally involve multiple entities of devices, cloud, and machine learning model providers, it is vulnerable to malicious attacks from adversaries against different entities.  Although existing work has made a lot of efforts to provide strong security protection for the FL systems, there is little work to defend or mitigate these malicious attacks from the system perspective. Bonawitz \textit{et al.} in \cite{ref-26} explored several more secure and robust aggregation algorithms and fault tolerance mechanisms from the perspective of system design. The serurity-enhanced techniques are designed from a system perspective so that FL can develop more practical industrial applications with the help of 6G networks.
\subsubsection{\textbf{Fair Federated Learning}}
FL involves thousands of devices training a shared global model in massive, heterogeneous networks \cite{ref-16}. Naive optimizing the global model in such a network may be unfair to some devices by causing disproportionate advantages or disadvantages. Obviously, FL towards fairness is an indispensable requirement for human-centric 6G communication services. Specifically, a fair FL in a wireless network involves fair resource allocation and a reasonable incentive mechanism. How to allocate computing and communication resources accurately and fairly in massive, heterogeneous networks has become a critical challenge that needs to be solved urgently. Some pioneering work, Li \textit{et al.} in \cite{ref-16} proposed q-Fair FL (q-FFL), which is a new aggregation algorithm to achieve a fair allocation of resources and accuracy. 
\subsubsection{\textbf{Explainable Federated Learning}}
The vast majority of FL models are black-box models (i.e., without interpretability), which makes users unable to understand what kind of services the model provides for themselves. In a complex 6G network system, the unexplainable predictions or decisions output by the black-box model may cause huge losses to users. For example, 6G-supported self-driving relies on an on-vehicle visual recognition model to determine whether the vehicle is running or stopped. Since the on-vehicle model has no interpretability, the driver cannot understand the decision of the vehicle model output. In 2018, the self-driving vehicle developed by Uber caused a car accident due to the wrong output of the on-vehicle black-box model \footnote{\href{http://tech.sina.com.cn/zt_d/uberincident/}{http://tech.sina.com.cn/zt_d/uberincident/}}. Therefore, in the context of a complex network system, such as 6G, the development of an interpretable FL model is the necessary way to human-centric communication services.
\subsection{Efficient and Effective Federated Learning}
\subsubsection{\textbf{Novel Asynchronous System}}
Even though the 6G network can bring the advantage of extremely low latency to the FL systems, the communication overhead is still the bottleneck of the FL systems being widely used \cite{ref-18}. As described in Section \ref{asy}, the two most commonly studied communication optimization schemes in distributed machine learning systems are the batch synchronous method and the asynchronous method (where the delay of the model update is assumed to be bounded) \cite{ref-26}. Indeed, asynchronous communication schemes involve scheduler, coordinator, worker, and updater, so there are several optimization problems for these roles that can be considered in the future: i) how the scheduler reasonably schedule the communication and computing resources in the systems; ii)  how the coordinator efficiently control the working state and idle state of the devices; iii) how workers and updaters optimize hyperparameters for model updates. These optimization problems are worth studying in future work in order to develop novel asynchronous systems.
\subsubsection{\textbf{Neural Architecture Search}}
The structure of the current FL models is generally predefined, but this predefined architecture may not be the best choice because it may not be suitable for non-independent and identical distribution (non-IID) data. Therefore, the Neural Architecture Search (NAS)-based Automating FL (AutoFL) schemes may be a promising solution to this problem. For example, a study in \cite{ref-27} proposed a federated NAS (FedNAS) algorithm to help distributed devices collaborate to find a better architecture with higher accuracy. NAS provides opportunities for seeking a better FL model architecture in the future.

\subsection{Towards Incentive  Federated Learning}
Existing studies mainly focus on enhancing the performance of FL algorithms, e.g., accuracy and training time. Nevertheless, an optimistic assumption, that all the data owners are willing to join the FL anytime and anywhere, is not practical in 6G scenarios with massive self-interest devices. As a result, incentive mechanisms for honest and active participation are a core and urgent research topic \cite{kang2019incentive,kangincentive,khan2019federated,weng2019deepchain,zhan2020learning,yu2020fairness}. 
Some interesting topics include: i) Due to information asymmetric between task publishers and participating devices, e.g., information about time-varying available resources, unfixed working periods and changeable participation willingness, it is still an open issue to design effective online-learning based incentive mechanisms to remove the impacts of both information asymmetric and time-varying factors, and also ensure efficient federated learning in 6G scenarios; ii) Considering heterogeneous and massive devices with diverse hardware equipment in 6G scenarios, the data quality of the devices is diverse. But the data quality plays an important role in learning performance. It is a challenging problem how to design data quality-based incentive mechanisms to motivate more devices with high-quality data to participate in federated learning and obtain higher rewards for their high-quality data contributions, thus improving both the system reliability and the learning performance \cite{kang2019incentive,khan2019federated,zhan2020learning}.
\subsection{Towards Personalized Federated Learning}
It is challenging for the FL system in the 6G network to provide users with personalized services. Prior studies \cite{ref-28,ref-29,ref-30,ref-31} adopt different personalized techniques to provide users with real-time personalized services, which is a solid step towards personalized FL. However, personalized FL still faces challenges from non-IID data, system heterogeneity, and network heterogeneity. Personalized service is a very important part of the human-centric 6G services. Therefore, it is an interesting and meaningful topic for FL to seek novel ways to address the above challenges.
\section{Conclusion}
In this article, we provided an overview of integrating federated learning into 6G communications. We discussed the requirements of 6G communication and core challenges of federated learning for 6G applications.  For the above challenges, we provided a comprehensive introduction of the emerging advanced federated learning methods for 6G communications, which including communication-efficient federated learning, secure federated learning, and effective federated learning.  Finally, we outlined out a handful of open problems and directions worth future research efforts. 



\ifCLASSOPTIONcaptionsoff
  \newpage
\fi

\bibliographystyle{IEEEtran}
\bibliography{reference}





\end{document}